# Mainstreaming Video Annotation Software for Critical Video Analysis

Matthew Martin, Auckland University of Technology, New Zealand
James Charlton, Auckland University of Technology, New Zealand
Andy M. Connor, Auckland University of Technology, New Zealand

*Abstract: The range of video annotation software currently available is set within commercially specialized professions, distributed via outdated sources or through online video hosting services. As video content becomes an increasingly significant tool for analysis, there is a demand for appropriate digital annotation techniques that offer equivalent functionality to tools used for annotation of text based literature sources. This paper argues for the importance of video annotating as an effective method for research that is as accessible as literature annotation is. Video annotation has been shown to trigger higher learning and engagement but research struggles to explain the absence of video annotation in contemporary structures of education practice. In both academic and informal settings the use of video playback as a meaningful tool of analysis is apparent, yet the availability of supplementary annotation software is not within obvious grasp or even prevalent in standardized computer software. Practical software tools produced by the researcher have demonstrated effective video annotation in a short development time. With software design programs available for rapid application creation, this paper also highlights the absence of a development community. This paper argues that video annotation is an accessible tool, not just for academic contexts, but also for wider practical video analysis applications, potentially becoming a mainstream learning tool. This paper thus presents a practical multimodal public approach to video research that potentially affords a deeper analysis of media content. This is supported by an in-depth consideration of the motivation for undertaking video annotation and a critical analysis of currently available tools.*

## Introduction

The analysis of video content is vital to quantitative and qualitative research in today's academic setting (Andersen et al. 2004, Hauptmann 2005, Roach et al. 2002, Wang and Parameswaran 2004). Researchers can acquire discrete or complex information from video sources through interactive methods and techniques. However, the extent to which researchers are aware of and value these techniques for information and data collecting in video content is still limited. Compared to manual, semi-automated and fully automated annotation methods within text and images (Erdmann et al. 2000, O'Donnell 2008, Stenetorp et al. 2012) it is evident to see the potential annotated videos can have for creating (multimodal) systems within research and education. Annotation of text is not a new approach and can trace its origins back to the addition of notes in the margin using pencils or pens. The emergence of digital formats and tablet-based technologies has reinvigorated text annotation and the use of natural language processing techniques (MacDonell, Min, and Connor 2005) allows the annotation of large text corpora to be virtually automated. Whilst more challenging, image based annotation is equally tractable in terms of automation (Yavlinsky, Schofield, and Rüger 2005) and even if undertaken manually the cognitive load in performing such annotation manually is relatively low due to the ease in viewing and analysing single images. As a result, text and image annotation is relatively accessible for the average computer user.

However, the knowledge and utilisation of annotating in videos by a typical applied researcher is still narrow and hindered from both the accessibility of related software and the trendy settings of basic use in amateur online videos. Virtually any video hosting website such as YouTube, allows users to layer external links and texts over top of a video without much effort (Khurana and Chandak 2013). Other platforms use approaches such as the creation of "tables of contents" for easy navigation (Li et al. 2000) or live annotating or 'chat' style tools only for active discourse between students, focusing on understanding already found knowledge instead of uncovering new information (Aubert, Prié, and Canellas 2014). These styles of annotation are popular with amateur users, but do not incorporate many other ways to annotate a video source to further enrich and understand the main content (Meixner et al. 2010, Qi et al. 2007, Wang et al. 2009).

Although for non-research purposes, commercial applications in film and television already illustrate effective forms of annotating-like effects (Taylor 2012). For one, in television the video content itself can include special effects to communicate specific or basic meanings to the audience with images or texts. Alternatively DVDs and Blu-rays include extra content with the main video such as subtitles, audio commentary, online linked resources, interactive scenes and so on (Bertellini and Reich 2010). This alongside others, are only internally known or accessed annotation methods by specialised video editors and filmmakers. They do however highlight how video annotation is already integrated amongst small communities. To do basic video annotating by researchers looking to adopt this approach requires a tedious process with specifically designed software, even for simple video edits. Whilst some researchers have attempted to address video annotation workflows (Hagedorn, Hailpern, and Karahalios 2008), there is still a high degree of complexity in existing approaches. On the other hand, with digital text content a user can easily copy, paste, highlight or refer a whole text or pieces of it to a web page, text document or even graphics editor. Here we discuss how these methods and basic functions can be well known and expanded to research analysis methods. Currently video hosting websites or related software are the limited choices for accessing video modifying tools, leaving many researchers with the only alternative to produce research outputs separately to the video source. If research analysis techniques are to evolve with updating technologies, an awareness of and access to improved tools is required for the average computer user. This paper proposes that there is the potential for video annotation (VA) to become a suitable addition to common video playback software found on computers, however to realise the potential there needs to be a wider adoption of the approach in research to drive a change in perception regarding annotation approaches. Annotations should be thought of as part of a video source rather than separate and alongside it. Video content is to be viewed as easily modifiable with the integration of annotations. Inspiration for this approach can be drawn from many of the public collaborations relating to annotating multimedia content (Carter et al. 2004). Just as literature content has become, we see VA requiring a diverse range of intended use and functions for it to result in common computer knowledge. To begin with, we have outlined many ways in which VA can function. The paper then illustrates where VA can currently be accessed and theories to why this is difficult. The focal goal for this paper however, is to identify VA as a simple form of video interaction for supporting research, and is exemplified in our own investigation of a software creation.

## Video Annotation

Video annotation is capable of being applied to a variety of media formats with a main video source, resulting in different ways for the VA to function depending on a person's interest. In this section we outline the many definitions of VA and how combinations of media formats can create different outcomes, mainly for research purposes. Many of the methods of using the media formats are less known to consumers and users of video content due to their subtlety in design and scarcity in availability.

The core function of VA should be to give users the ability to interact with content and offer an active mode of utilisation. With even minor forms of interaction found in video playback applications a user can slide between video content to create a non-linear flow (Yeung and Leo 1996). Taking this further, slicing up video content allows a user to compare between different shots, or limit the viewed content to specific actions or objects. Just like literature search functions can skip sections of papers by locating chapter titles or specific words, video software capable of identifying specific objects can have a user focus on one topic or action. Whilst this approach is already achieved through the creation of tables of contents that are associated with a video, these approaches normally are constructed separate from the video itself. A more interesting approach is to embed the data into the video to enhance the shareability of the rich data. It has already been noted by other researchers that most existing VA tools lack the ability to integrate a variety of data sources in any unified way (Rich and Hannafin 2009).

In terms of automation of the VA process, software algorithms have developed to have the capability of identifying simple objects within video content. Probably the simplest example is the identification of actual text in a video (Zhang and Kasturi 2008) though it is also possible to identify objects (Sivic and Zisserman 2003). For example, searching for a 'foot' object in an athlete's sports performance video could identify and compare frames specific to understanding correct footwork techniques based on the automatic extraction of the frames. Whilst such automation is of interest, there is a need to ensure that the basic VA workflow is correctly setup to ensure that the basic process can add value to the users prior to attempting more complex tasks. For video interaction to function effectively it should not be disruptive for the user (Aubert, Prié, and Canellas 2014). If the interactions are too strong on the user as they try to consume the content they may lose their train of thought. Effective annotation interaction for video can be suggestive or hint towards an interaction while users indulge in the video content.

A common function of VA is layering text on video. It is easy to produce and highly changeable (Aubert, Prié, and Canellas 2014). As subtitles in films, visually it communicates sounds and words from a video for a hearing-impaired audience, providing access to information not reachable within the original content. Alternatively, using audio for annotating words displayed on screen can work as an educational tool for correct pronunciation (Aubert, Prié, and Canellas 2014). This too can help with accelerating user learning of video material. Researchers who face similar issues with recorded material (hearing impairment or pronunciation difficulties) could consider this a useful function to aid their analysis. However as a basic reflective process, text based annotation for video analysis can be found in websites like YouTube. Here the video uploader can use annotations for reflective communication to the viewers, giving mention of mistakes made or parts found interesting in the video. What may not be entirely expected by the uploader is that while they consider how to embed annotations into the video they have already begun a process of analysing the content.

An annotation can assist in building relationships between original content and related material via hyperlinks (Wolfe 2002). Already this is found amongst academic papers, providing hyperlink directories of cited references and leading readers to specific areas of a paper or external website. As video essays become a popular hand-in component for students, it too can be designed with hyperlinks to cited material. A researcher can therefore make new paths for resource gathering if they are able to navigate information freely.

In terms of analysing the video content a user is most likely to inspect it in smaller pieces. Already there are programs capable of highlighting a person or object in moving imagery (Vondrick, Patterson, and Ramanan 2013), and indeed once located a moving object can be automatically tracked in subsequent frames (Kim and Hwang 2002). As a result, in an overwhelmed area of visual content a user can block out or focus on a certain image through automated as well as graphical techniques, creating a new perspective for what might be happening in the content (Aubert, Prié, and Canellas 2014). This is to be considered as hiding the many layers to the video resource. With annotation tools used to visually block content in a video, it has potential to filter out less important pieces of information.

Another method of analysis through VA includes data visualisation, representation or summarisation. Information found in a video can be translated into new types of data or condensed down, making the conclusive results more obvious (Khurana and Chandak 2013, Ma et al. 2002, Vu et al. 2013). Not only is this approach useful in terms of providing more accessibility to the content of the video, but the new data can also be compared again with the original content to find trends. An example would be counting the amount of pedestrians walking down a road and graphing them into categorical variables. With these graphs the user can have direct correlation with the video to reveal why specific groups of people walk down the road.

This generates a looping feedback for the research, pulling the data out and recirculating it back into the content. Alternatively the data can be given semantic value to gain user attention quickly, reminding of what the data was when relooking at old data (Snoek, Worring, and Smeulders 2005).

Visual data for observational learning is less direct in its definition than literature. It is not as literal to its meaning. Therefore visual work is interpretative and allows for researchers and students to discuss the content and collectively moderate new knowledge (Aubert et al., 2014). Similar to sharing opinions or interpretations of video content with others, a user can communicate a thought or concept to him or herself throughout the analysis process of the video. Here by simply placing a few lines of text effectively on top of the video it can identify new connections for the researcher. The timing of a text-based message creates a simplified way of defining what is happening in the visual content (Golan et al. 2002).

The different methods to using VA make it apparent that there are ways to use it to form new questions and new connections between information. The type of annotation function the user is to incorporate will enable different information to be revealed. Annotation is usually known for its assistance in remembering or communicating information across different time periods and platforms, but it can also become a tool for digging into already obtained, hidden data. A way to consider it is as a means of discovery. Annotating video content is then to be considered as a tool to searching for valuable and questionable outcomes out of a large volume of otherwise inadequate material.

## Accessibility

The limited accessibility of VA software is a large influence to why it is not a well-established research tool. While it is possible to discover many online types of VA software, there are a number of barriers preventing a researcher from utilising them, many relating to skill level or price. This section outlines some of the software a person could use for VA and theories of why it is not easily accessible. We relate a lot of these problems to the reasons for why academics, communities, and commercial companies do not have a sense of importance for developing such software.

There are various types of video annotation already in existence, all with different features and uses. Previous surveys have provided a list of available tools for VA (Aubert, Prié, and Canellas 2014). Most are listed with a specific audience in mind and relate to the user's job. This entails a narrow list of features, setting aside the general, mainstream uses like importing, exporting and the variety of interaction and sharing with the videos. For instance, one tool mentioned called VideoTraces was designed and used for learning in recorded dance sessions. Other tools like CLAS or MediaNotes or MOOC courses are designed specifically for students and stay within a student based audience. Here lies the problem. The designer already anticipates the expected use and type of user for the software. This leaves the user with a limited amount of function and outcome. Looking back at VA and literature annotation tools illustrated in Joanna Wolfe's Annotation Technologies such as Debora, MRAS or CommonSpace (Wolfe 2002) are now outdated and neglected. Similarly they were designed within a small context for a certain target audience instead of focusing on the core video content.

The video format is always evolving and as such it is accustomed to constant updates and changes. Any video playing software already requires revisions and updates to work with new firmware, file formats and integrated features. Adding in enhancements like separate audio, text, web integration etc. creates further complications for the large volume of information included in video files. Having such complications would imply new systems around video technology for the public to understand, creating costs that exceed a developer's current basic and freely available video playback software. Companies are unlikely to go to such efforts for this software without seeing a reasonable profit (Wolfe 2002). Another understandable reason for current video playing software not supporting annotations is it is virtually only capable of reading content and not writing (editing) it. To do anything to the video or be held responsible for the annotation may create intellectual property issues with both parties' (Aubert, Prié, and Canellas 2014, Wolfe 2002).

Online communities utilising video manipulation have existed in independent, commercial and learning areas (Brown and Adler 2008). This has allowed for video manipulation tools to be available using open sourced platforms. Surprisingly, it is difficult to find open source software with direction towards individual academic research analysis. Many open source software target user communication between viewers of the video through online hosting. For instance, one web based library called Open Video Annotation Project (OVA) at Harvard University confines within online video hosting, allowing researchers to display annotations as notes to each other. The annotations can be locally exported to have further analysis and the library code is available for access, but the tool is limited to use as a web browser plugin. As many of these types of software all have similar video hosting traits, a barrier faced by developers could be related to intellectual property of these videos.

Software designed for academic annotation purposes such as OVA, are typically limited to support from within one's own research (Wolfe 2002). This support is never long lasting as the software's usability survives as long as the research, never reaching a sustainable and widespread audience. This can be from disputes between general academics and researchers not seeing the full potential of annotating, specifically within videos, as a worthwhile tool for research. Affiliation with traditional literature papers and paper referencing are likely reasons for not being as open to new techniques. Also as video is less concrete as a reference due to its temporal properties making it difficult for exact descriptions, the use of videos along with other digital media in academic papers has not yet gained popularity (Pearce et al. 2010) Despite this concepts of electronic journals and embedded media have been discussed for some time (Kling and Covi 1995). Therefore, outside of VA being related to an academic's research, it has never been completely utilized for its expected use.

From a researcher's perspective, video is far from being as influential and adjustable as literature based work. It does not have an effective means for customisation in popular software and restricts developers legally from being held accountable to giving a user access to such tools. There are many extensive research outcomes on VA tools benefiting learning and research, evidently producing valuable tools for VA research analysis. However, the lifespan and audience of these tools continue to stay underdeveloped and without sustainable results for users. Providing anyone access to open source VA tools amongst general communities and platforms is a likely progression to changing how users and academics see and interact with video software. It can also lead to stronger implementations as programmers develop the open software amongst each other.

The following section outlines the implementation of a prototype tool for video annotation to be utilised by researchers in the first instance. Unlike the Open Video Annotation Project, this tool can be used on locally stored content. The intention of the tool is to be an open source project that attracts a wide range of contributions who will develop the tool through a number of iterations in order to both grow the included functionality and both identify and address issues regarding usability. We hope that developing a community of video annotation researchers will assist in defining possibly applications for video annotation outside of the research community.

## Implementation

To fully realise many of the arguments made throughout this paper a prototype of our own annotation tool was created. It was implemented in support of an ongoing research project, including video as a core research analysis component, however has been adopted by a number of other researchers. The objective of developing the prototype was to get a better understanding to the limitations of local VA tools and the process to creating a useable version. This informed us to how feasible a VA tool creation is in relation to a programmer's skill level and time spent in its development.

The final outcome of the tool is aimed for playing high quality videos with basic annotation interaction for anyone to be able to use. For us to understand a VA workflow we kept to text only overlays over the video with manual object tracking with the mouse. The tool is to work as a standalone application, easily transferrable between computers.

The software was created using freely open sourced programming tools found on a Mac operating system. Without going into much detail on the technicalities it should be known that the software uses the QuickTime player with the help of Perian software for its video playback. The timeframe of the software's development spanned just a few weeks, with extras functions like slowing video pace and autosave implemented to improve the software's functionality. To begin the program the user selects a video file on their machine as shown in Figure 1.

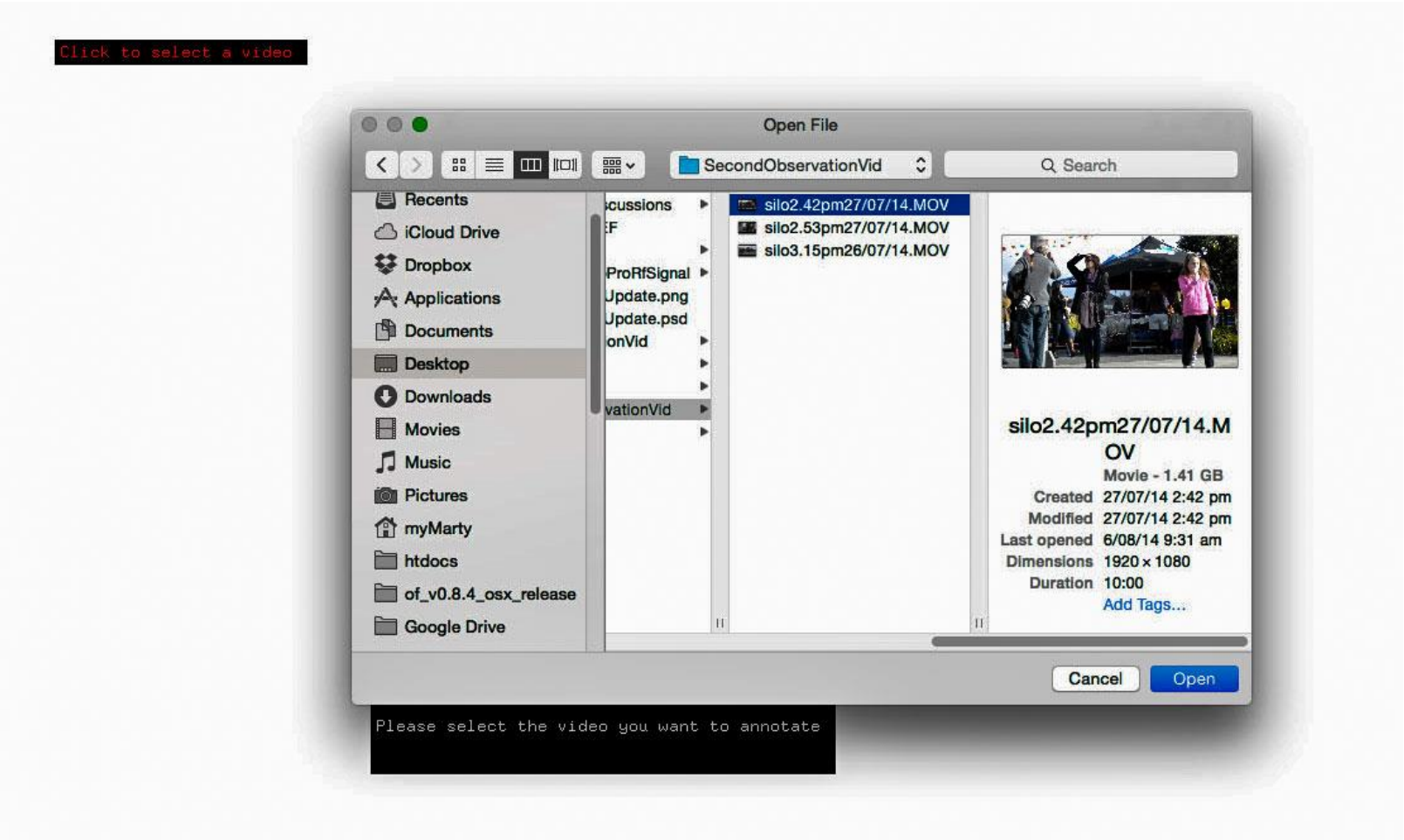

Figure 1: The User Selects the Video to Annotate

The user is then asked if they want to select any previously saved annotations onto the video. The annotations are currently encoded in a txt file, stored in the application's internal folder. By separating the annotations from the video allows users to integrate different annotations as they please, creating new ways of comparing data. However, this will most likely cause technical errors, as each annotation file will have different properties from the duration and dimensions of a video. Therefore the annotations would either be inclusive of the video or contain forced restrictions relating to the video. We implemented the latter. After the video is loaded the user can interact in various ways. Common video playback options are included, like moving the current position of the video or changing volume. Creating an annotation is through a simple button press followed by typing in text. Replaying or rewinding the video will show the annotations in their positions and time as they were instructed. Annotations can be created while the video continues to play or can be paused to focus on a single frame.

The use of our tool is illustrated here in terms of current research by the authors and the functions available of the tool. It is a recommended scenario of how a user could interact and benefit from the tool based on video annotation outcomes produced by the authors' research.

The user of the tool begins with the intentions of analysing and identifying connections between objects and people in public space. The annotations simplify each object to a word or short description, highlighting things like a person's interaction with an object. The user watches through the video to get an initial comprehension, only adding a few annotations to basic objects. On the second watch a higher influx of annotations are to fill the space, creating a larger focus on the less noticeable number of objects. By the third playthrough the user begins to find patterns between the annotations as the annotated objects appear frequently in the public space. This is illustrated in Figure 2.

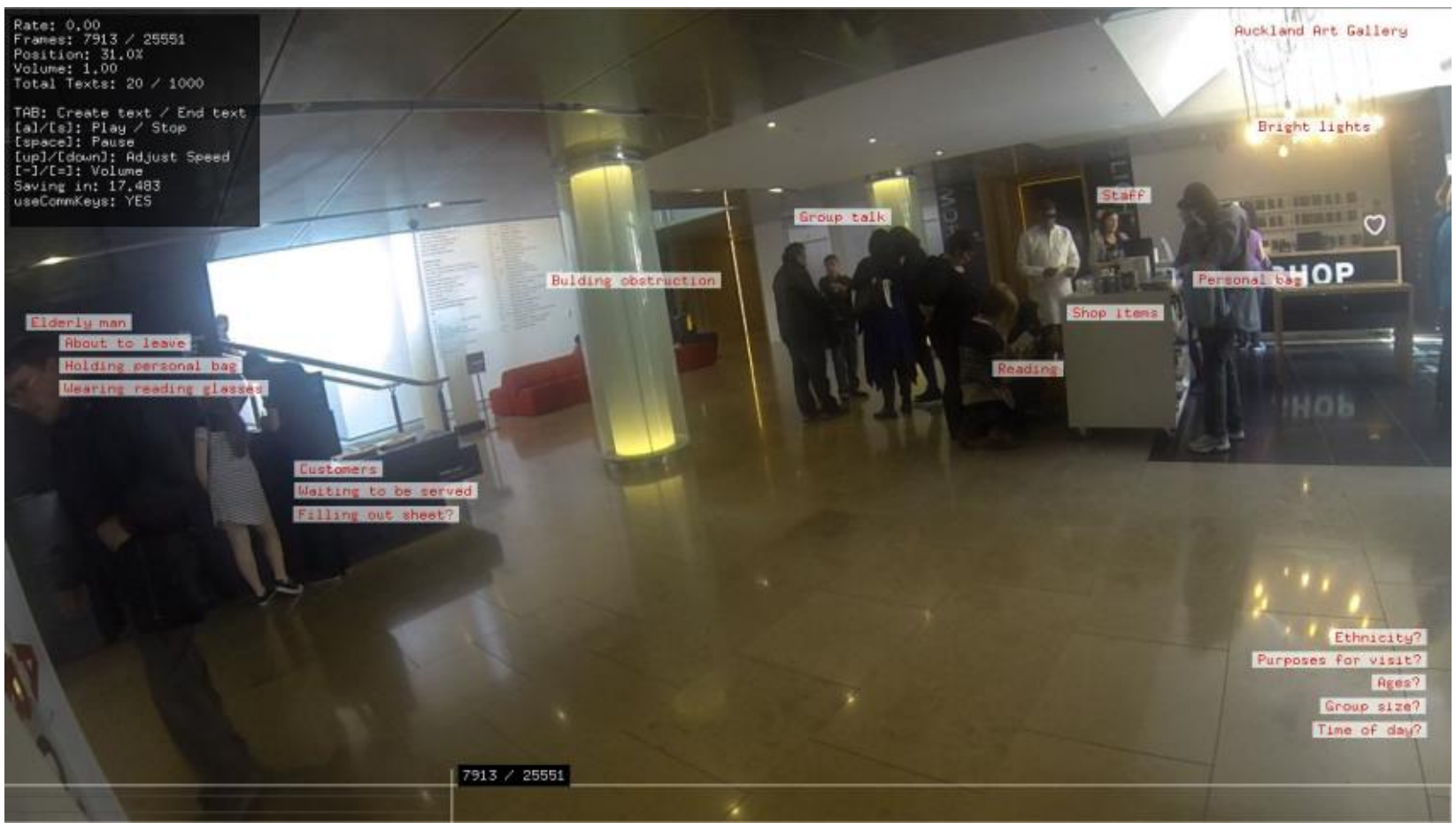

Figure 2: Text Annotations Identifying Objects and Behaviours

By using our tool a user can have basic abstracted text on top of the video, generating different thoughts and focuses for the researcher. This brings initial questions to the annotated object and its importance in the context of the video. Such questions would be skimmed over if done through less direct noting of the content. These objects are also brought into a new category (the annotated category), linking them together. For example, as shown in Figure 3, if a pedestrian using a mobile phone walks by while holding hands with their partner in the video, the researcher would annotate each item. On second play through the user begins to identify a relationship between the three as they were annotated. This might lead to questions on the impacts a phone brings to a user's environment awareness and their methods to walk without looking (i.e. use partner as a guide).

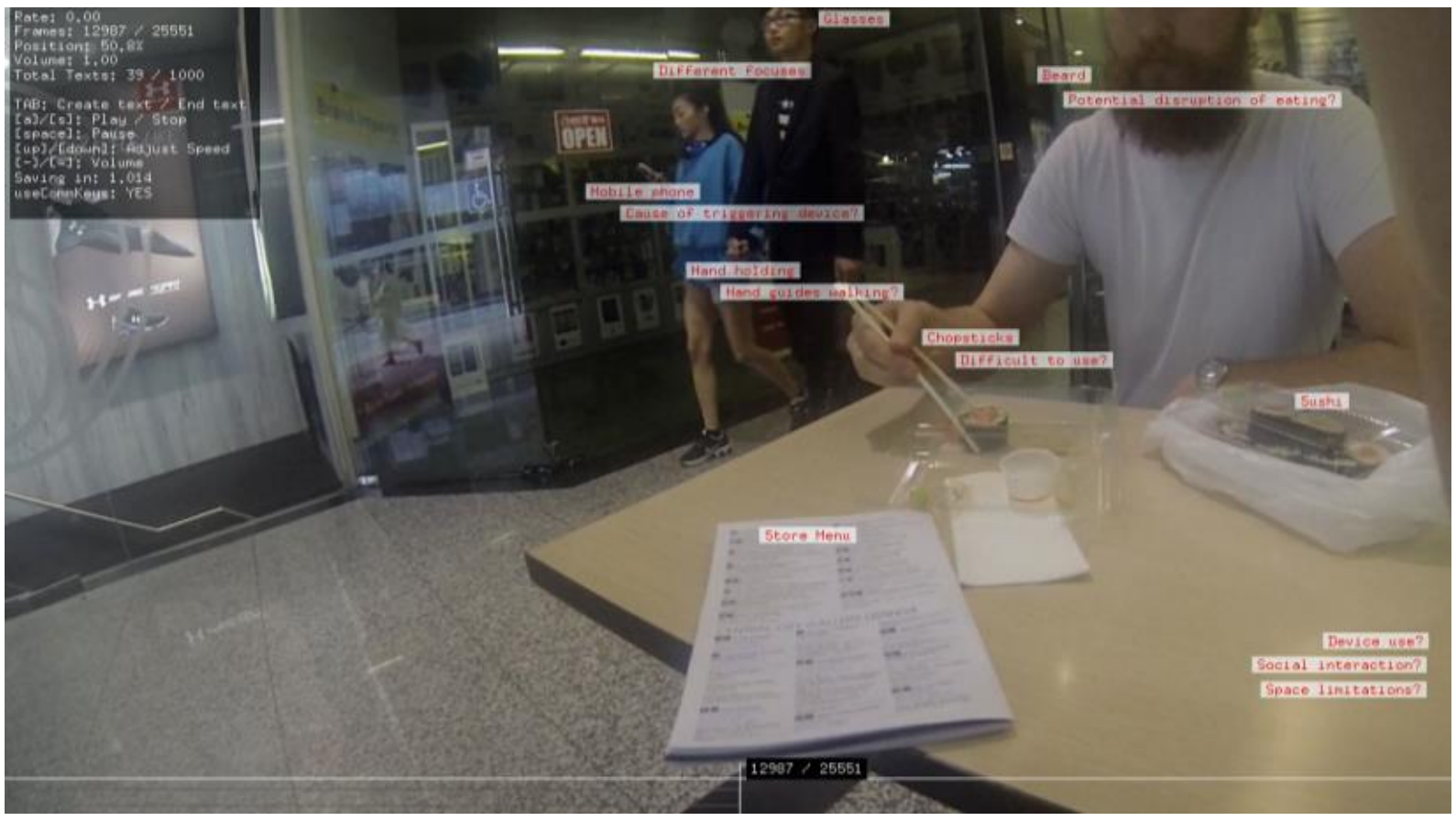

Figure 3: Capturing Relationships with Annotations

After a prolonged time of adding in annotations the impact can fade. This can additionally bring the non-annotated objects to the foreground. We consider this a delayering of the different objects and information of the video. The use of annotating can also be overly done. Constant highlighting of objects in a single play through can result in confusion between different moving texts on replay. To keep to an appropriate amount of annotating it is up to the user to form their analysis goals through appropriate portions of annotating. The tool is therefore dependant on the user correctly annotating to some degree as part of the analysing process.

From the short time making the tool along with the promising research outputs from basic annotation, we consider the development of these tools to be valuable in relation to the effort. Even with simple function VA can create justifiable results for researchers to progress with. The focus is then for the possibility of such software being available to computer users. From the amount of time and effort put into the software's creation it is plausible to see such programs available as a standard tool on personal computers.

## Future Embedding

Despite the limited uptake to date, there are many more possibilities that can be envisaged for developing VA as part of mainstream tools. Such development requires an understanding of how VA can be used in practice to ensure the tools support a range of appropriate workflows. Whether for a commercial product or openly available through online means, an important step forward for VA is its recognition as a valuable application for anyone to access. Such an understanding will be developed through the ongoing research projects that utilise the current VA prototype as well opening up the tool to a range of additional developers and users through open source licencing.

Here we have suggested approaches to benefit VA becoming more widespread in research, potentially even becoming a more open and mainstream activity in its own right. In multiple ways our suggestions are based on text software and its emergent of complex, multifunctional systems for literature annotation. Initially we see VA implemented as part of current mainstream platforms such as VLC and QuickTime. This will help generate understanding of how VA can be applied, as videos shared amongst people will begin to contain the annotations. Keeping focus on research outcomes, we then discuss the importance to creating suggestive features into the platforms to direct the user towards new learning patterns. These features are based on two specific video annotation tools that were not designed for a general audience. Later, it is discussed where VA can be applied in future video recording or playback technology to maintain its support.

As covered, the use of VA is difficult to come by for anyone not equipped with video editing expertise. This to some degree has been abolished from our own quick implementation of VA, showing it can be produced and released within a user's reach. However, making annotation software, even if it is fast to produce, will not mean that it will be well used by computer users. Keeping to an already established audience would increase the chance for anyone to know about its presence as YouTube has already proven with their video hosting annotations. For this reason it is suggested using established video playback programs would be a stronger supporting option if VA were to be embedded in a mainstream environment. Popular video playback software like VLC and QuickTime could then be reframed as not just video reading tools but also video writable ones. Promoting users with such options would not just allow the software to integrate further editing tools but also what users can anticipate doing with the video itself. The prototype created as part of this paper is a type of extension or plugin to the QuickTime player and could potentially work as part of the core QuickTime application.

Text based content on computers has been readable and writable between different software for decades. The many options encompassed in word processors like Microsoft Word and OpenOffice give the user many preferences to manipulate text data, with the ability to copy over changes between software. Having multipurpose options for text data between software makes those manipulations common knowledge for users and allows people to recognise the software as an annotating tool. If videos were to be supported as changeable and transferrable between software like VLC and QuickTime, the applied use of annotating videos would also be known as accessible with the software. Videos could then have associated annotation files like images have XMP files for storing metadata information. If videos were able to contain additional information through extended files or an entire new format, writable video content would begin to become an industry standard.

As VA progresses into becoming part of mainstream software, incorporating guides and structures can help to articulate VA's effective use in education and research (Golan et al. 2002). Software such as Moodle and Animal Landlord has proven to shift students learning and education practices in communication and analytical methods. Moodle gives students options to communicate freely in online courses amongst each other (Manenova and Tauchmanova 2011). They can send text or audio messages from one another, as well as to the teacher, and place this discourse into the visual content of the online class.

Older software called Animal Landlord was designed in 2002 to focus on pushing high school students into effective analytical methods of video research and change the educational learning process. Animal Landlord was an analysis tool specifically created for deconstructing the behaviour patterns of animals (Golan et al. 2002). It gave high school students a choice between limited options to describe what an animal was doing. When the student had decided, the description was placed in either an observation or an interpretation category. Limiting the choices like this for the user meant certain analytical methods had to be followed to understand the content.

Providing similar options to mainstream software would invite new ways for users to think about their videos. In simple terms it could be to label the content into genres or themes. More detailed uses for example, could be to describe what the viewer enjoyed about a fictional film or select their most likeable character for certain scenes. These examples are for the general video audience but could be specified towards research methods within academic presentations or tutorial videos, asking viewers to highlight relevant sections in relation to their research.

In order for VA to succeed as a desirable method of learning and research it needs to impose, ever so subtly, new thinking patterns for ways to interact and learn from the video content. The software that incorporates VA should still be open to multiple uses, however it should include guidance for analytical purposes. If methods from Animal Landlord and Moodle were to be implemented into video playback systems, it would provide another step towards developing VA as an everyday learning tool.

If students have questions or want to further their studies with extra curriculum in areas not completely covered in the video content, the use of VA would lead the user to related content by giving the user selectable options. For example if the user wanted to find out more info on a historical person in a history video, they would highlight them and the software would display paths for the user to follow down; like biography, timelines or family relations. Of course this could be examined in many faculties and even allow for cross-referencing between different topics. These options are up to the user as to how they want to learn but can benefit their education in narrowing focus or widening curriculum and cross referencing topics.

We see these techniques as something to be highly effective amongst mobile technology, especially for younger generations as they grow up with these systems highly visual based tools. Through the use of touch screens, selecting and highlighting in video content can be done via the user's fingers, creating an understandable relation between computer and user. This type of technology is expected to be highly interactive because of its touch screen controls and a promising starting point for VA being an actively interactive tool. Also, because tablets are already accustomed to most modern schools today the VA would be an embraceable addition.

So far the proposed implementations for embedding VA have revolved around existing technology. However, as technology evolves so too can VA. An example for this would be the emergence of wearable devices becoming common in public spaces (Starner 2013). With portable cameras people can take videos on the go, gathering information as they move through their day which has allowed so-called "life-logging" to transition from a quirky curiosity to a major challenge in terms of dealing with big data (Gurrin, Smeaton, and Doherty 2014). The continuously recording device gives opportunity to develop autonomous analysis of visual content as it happens. Already video recording programs can capture specific information, such as the recordings location, human face tracking or sign and object identification (Brilakis, Park, and Jog 2011, Eleftheriadis and Jacquin 1995). This information can be fed back to the user, through the device's output system and influence what decisions the user is to make. Users can expect an active space for highlighting, linking, communicating aspects of their recorded life, continuously using this data to be informative. This again is a type of annotation that can be expected to be part of future devices like the wearable recording ones.

Our propositions for VA are a small dose of how it could be applied for later use. There are many other techniques to which VA could be implemented but we consider these important steps towards a mainstream outcome of VA, specifically with research analysis integration. To get anywhere however, first there needs to be a starting point for video content to be known as an interactive material. If VA options become available to video users on their computers they can change their thinking and engagement with the video to a more active mode. The awareness to such engagement with videos will lead to the communities within particular areas, including learning and research. If VA becomes a habitual tool, further demand of it within video software will provide competent development into emerging mainstream technology.

## Conclusion

From the given examples stated throughout this paper it is clear to see a couple of differing intentions and contexts for annotation and modification of videos. Commercially, annotations are designed within professional software, for users with the knowledge of and means to buy such software. A range of techniques and methods for VA are capable and effective with the software, but out of reach for a standard computer user. Less supported are developers with academic backgrounds contributing to open and accessible tools, but are confined to the audience and goals of their research. As expected it has led to outdated and inaccessible outcomes. We also see from the online video hosting and sharing means like YouTube and MOOC courses it is possible to have unqualified users annotate freely, but again this is confined to the very basic VA methods by the video hosts. Any mainstream local video reading software is therefore what we consider the strongest approach to support sustainable interactive and writable tools for videos but may be dependent on legal restrictions of video content.

Changes to these restrictions will occur as the format becomes modifiable between different tools and users, potentially becoming an industry standard VA format. As we have already illustrated ourselves, it does not require much effort to produce a personal program with these capabilities, and with enough interest between academics, developers and other video editing communities it can become a sustainable and popular tool amongst the public and their computers. At this stage the development of the tool has been completed, but to date the usability of the tool has not been formally evaluated. The tool is in the process of being made available as an open source project and this will lead to the collection of data regarding usability and desired functionality. The knowledge of the tool as a commodity on computers will lead to its success in research and research analysis. It is obvious to see the potential of VA and with support from established tools and communities, the everyday use of video annotation will move closer to the already accustom, literature annotation.

## ABOUT THE AUTHORS

***Matthew Martin:*** Auckland University of Technology. Matthew obtained the Bachelor of Creative Technologies degree in 2013 from Auckland University of Technology in New Zealand. He is currently working towards his Master of Creative Technologies degree at Auckland University of Technology (AUT) in Auckland, New Zealand. His research interests relate to the role of technology in terms of generating experience.

***James Charlton:*** Auckland University of Technology. James is an artist and a Senior Lecturer at Colab. He engages a range of physical, digital and performative approaches to explore the nature of the artefact and the assumptions of the audience. James gained his BFA from Elam School of Fine Arts in 1982. As a Fulbright recipient he completed his MFA at the State University of New York in 1986, and exhibited extensively throughout the United States. James lectured in Sculpture at the University of New Hampshire, Monserrat College of Art and the State University of New York at Albany. His current research is in the area of interactive digital object technologies centres around the integration of digital and physical content to question the definitions and nature of time-based media.

***Andy M. Connor:*** Auckland University of Technology. Andy is a Senior Lecturer at Colab, the "collaboratory" at Auckland University of Technology in New Zealand. His undergraduate training is in mechanical engineering and he holds a PhD in mechatronics. He has worked at the Engineering Design Centres at both the University of Bath and the University of Cambridge in the UK. Following a number of years of commercial experience as a software engineer and a systems engineering consultant, Andy migrated to New Zealand and took up a number of roles in software engineering and computer science at Auckland University of Technology prior to joining Colab in 2012. Andy has a broad range of research interests that include automated design, computational creativity, education, evolutionary computation, machine learning and software engineering.